\documentclass[final,5p,times,twocolumn]{elsarticle} 
\usepackage{amssymb}

\usepackage[table,dvipsnames,x11names]{xcolor}
\usepackage[table,dvipsnames,x11names,addedmarkup=bf,addedcolor=BrickRed,deletedmarkup=sout,deletedcolor=Blue4]{changes}
\usepackage{amsmath}
\usepackage{multirow}
\usepackage{booktabs}
\usepackage{bm}
\usepackage{braket}
\usepackage{amsmath}
\usepackage{graphicx}
\usepackage[normalem]{ulem}
\biboptions{sort&compress}

\definechangesauthor[name={Xie Mengran}]{xmr}
\definechangesauthor[name={IMP}]{redauthor}
\definechangesauthor[name={Blue Author}]{blueauthor}

\usepackage[pdfstartview=FitH, colorlinks,
 linkcolor={red!60!black!},
 anchorcolor={green!80!black!},
 urlcolor={blue!80!black!},
 citecolor={blue!50!black!}]{hyperref}

\journal{Physics Letters B}

\begin{document}

\begin{frontmatter}

\title{The role of the overlap function in describing angular distributions of single-nucleon transfer reactions}

\author[ad1,ad2]{M. R. Xie}
\author[ad1,ad2,ad3]{J.G. Li\corref{correspondence1}}
\author[ad4]{N. Keeley\corref{correspondence2}}
\author[ad1,ad2]{N. Michel}
\author[ad1,ad2]{W. Zuo}

\address[ad1]{State Key Laboratory of Heavy Ion Science and Technology, Institute of Modern Physics, Chinese Academy of Sciences, Lanzhou 730000, China}

\address[ad2]{School of Nuclear Science and Technology, University of Chinese Academy of Sciences, Beijing 100049, China}

\address[ad3]{Southern Center for Nuclear-Science Theory (SCNT), Institute of Modern Physics, Chinese Academy of Sciences, Huizhou 516000, China}

\address[ad4]{National Centre for Nuclear Research, ul. Andrzeja Sołtana 7, 05-400 Otwock, Poland}

\cortext[correspondence1]{Corresponding author. e-mail address: jianguo\_li@impcas.ac.cn (J.G. Li)}

\cortext[correspondence2]{Corresponding author. e-mail address: nicholas.keeley@ncbj.gov.pl (N. Keeley)}

\begin{abstract}
Single-nucleon transfer reactions offer a valuable way to probe nuclear structure. We explore the effect of directly introducing overlap functions computed using the Gamow shell model (GSM) into reaction calculations, taking the $\left< ^7\mathrm{Li} \mid \protect{^6\mathrm{He}} + p \right>$ single proton overlap as a case study. By incorporating both inter-nucleon correlations and continuum coupling, the GSM  provides accurate overlap functions in both interior and asymptotic regions, together with the corresponding spectroscopic factors (SFs). These theoretical SFs and overlap functions were included in a coupled channels Born approximation analysis of the \(^{6}{\rm He}(d,n)^7{\rm Li}\) transfer reaction. Overlap functions derived from \textit{ab initio} no-core shell model (NCSM) calculations as well as standard single-particle (s.p.) wave functions were also considered for comparison. Our results reveal significant differences between the calculated angular distributions when employing theoretical SFs with standard s.p.\ wave functions compared to the full theoretical overlap functions. Discrepancies were also observed between angular distributions calculated with GSM and NCSM overlap functions, highlighting the importance of internal structure and correct asymptotic behavior in reliable reaction calculations. The GSM overlap functions also provided a good description of the $^{208}$Pb($^7$Li,$^6$He)$^{209}$Bi reaction when included in a coupled reaction channels calculation.
\end{abstract}


\end{frontmatter}

\section{Introduction}
The advent of radioactive ion beams has enabled the spectroscopic investigation of exotic nuclei~\cite{Gade_2016, AUMANN20073, BORGE2016408, Gales_2011, Andrighetto_2018, brown2024motivationsearlyhighprofilefrib, Sakurai2010}. Single-nucleon transfer reactions, such as $(d,p)$, $(d,n)$, $(d,t)$, and $(d,^3 \rm{He})$, have proven instrumental in providing insights into the structure of both target and residual nuclei, particularly in the determination of single particle (s.p.) levels~\cite{TIMOFEYUK2020103738, doi_10_1142_5612, PhysRev.80.1095, PhysRev.80.1095.2, PhysRev.85.1045.2, RevModPhys.32.567} and  spectroscopic factors (SFs)~\cite{PhysRevLett.21.301, PhysRevLett.104.112701, PhysRevLett.131.212503, PhysRevLett.102.062501, PhysRevC.73.044608, PhysRevC.78.041302, Wimmer_2018, PhysRevC.94.024619, PhysRevLett.19.90}. 
In transfer reaction theory, the cross section is typically expressed as the product of a theoretical s.p.\ cross section and a SF~\cite{AUMANN2021103847}:
\begin{equation}
    (\frac{d\sigma}{d\Omega}) \sim C^2S \cdot (\frac{d\sigma}{d\Omega})_{sp},
    \label{E1}
\end{equation}
where $C^2S$ denotes the SF, including the isospin Clebsch–Gordan coefficient $C$ and $(\frac{d\sigma}{d\Omega})_{sp}$ the cross section calculated with a reaction model, e.g.\ the distorted wave Born approximation (DWBA), using s.p.\ wave functions~\cite{Li2010, AUMANN2021103847} rather than overlap functions, known as the s.p.\ approximation.
The above framework is employed to  extract SFs from experimental single-nucleon
transfer reaction cross sections.


Due to the frequently predominantly peripheral nature of the reaction, single-nucleon transfer cross sections exhibit a sensitivity to the asymptotic behavior of the overlap function, which has the form~\cite{TIMOFEYUK2020103738}:
$O_{\ell j}(r) \sim W_{-\eta,\ell + 1/2}(2\kappa r)/r$,
where $W$ is the Whittaker function, $\kappa = \sqrt{2\mu S_n}/\hbar$, and $S_n$ the  separation energy of the transferred nucleon. When using s.p.\ wave functions for the overlaps, the depth of the Woods-Saxon (WS) binding potential is adjusted to match $S_n$ to give the correct asymptotic behavior~\cite{AUMANN2021103847, TIMOFEYUK2020103738}. However, while this procedure ensures a proper description of the tail region, transfer cross sections are also sensitive to the radial shape of the overlap function~\cite{PhysRevC.97.034601}, which significantly impacts both the shape and magnitude of the angular distribution. Consequently, the s.p.\ SFs are strongly dependent on the binding potential geometry, leading to variations of up to 30\%, even with reasonable parameter choices~\cite{TIMOFEYUK2020103738, MA201926, AUMANN2021103847}.

For transfer reactions involving weakly bound or resonance states, the situation becomes particularly challenging. Such systems typically exhibit extended spatial distributions and enhanced continuum coupling, making the calculated angular distributions highly sensitive to both the interior and asymptotic parts of the overlap functions—features that cannot be accurately described by s.p.\ overlap functions.
Addressing these challenges remains a significant hurdle for current direct reaction theories, emphasizing the need for more precise nuclear structure inputs.
Moreover, the theoretical calculation of overlap functions, especially their asymptotic behavior, also provides great challenges for nuclear structure models.
Among these, the Gamow shell model (GSM) stands out for its ability to account for both nucleon correlations and continuum coupling, enabling an accurate description of nucleon wave functions, especially in the asymptotic region~\cite{XIE2023137800, Xie2023}. With the GSM, reliable overlap functions and SFs can be obtained as inputs for reaction calculations.


The proton SF of $^7$Li has long been debated~\cite{COHEN19671, PhysRevC.16.31, PhysRevLett.82.4404, Li2010}. Meanwhile, $^6$He, as a two-neutron halo nucleus, poses additional theoretical challenges due to its extended spatial distribution and weak binding. These structure features make the $^6$He$(d,n)^7$Li transfer reaction—where angular distributions have been measured~\cite{Li2010}—an ideal testing ground for investigating the role of SFs and overlap functions in reaction calculations.
In this work, the experimental data are reanalyzed using both GSM overlap functions and GSM SFs in conjunction with 
standard s.p.\ wave functions. Direct use of the GSM overlaps provides a better description of the shape of the measured angular distribution, as well as having a significant impact on the magnitude of the cross section. 

\section{Method}

\subsection{The nuclear structure model: Gamow shell model}
The GSM employs the one-body Berggren basis~\cite{BERGGREN1968265}, which possesses bound, resonance, and scattering states. In the GSM a nucleus is described as valence nucleons interacting outside a closed $^4$He core, and the valence nucleon coordinates are defined with respect to the core center-of-mass (c.m.) in the Cluster Orbital Shell Model (COSM) framework~\cite{PhysRevC.38.410}. The GSM Hamiltonian consists of a one-body core-valence potential, a two-body interaction between the valence nucleons, and a recoil term~\cite{PhysRevLett.89.042501, PhysRevLett.89.042502, Michel_2009}. The core-valence potential is modeled by a WS potential, comprising central, spin-orbit, and Coulomb terms, while the two-body interaction is described by the Furutani-Horiuchi-Tamagaki (FHT) effective nucleon-nucleon force, possessing central, spin-orbit, and tensor parts with Gaussian radial dependencies~\cite{10.1143/PTP.62.981}. The Coulomb interaction between valence protons is treated exactly. The parameters of the GSM WS+FHT Hamiltonian were fitted to reproduce the features of light nuclei, as detailed in Ref.~\cite{Jaganathen:2017}. 

Within the GSM framework, overlap functions are defined as~\cite{PhysRevC.75.031301, Michel_Springer}:
\begin{equation}
O_{\ell j}(r) = {\frac{1}{\sqrt{2J_A+1}}} \sum_n \braket{\Psi^{J_{A}}_{A} || a^+_{n \ell j} || \Psi^{J_{A-1}}_{A-1} } u_{n}^{(\ell j)}(r), 
\label{E3}
\end{equation}
where $\ell$ and $j$ denote the orbital and total angular momentum of the partial wave, respectively. $J_{A-1}$ and $J_A$ represent the total angular momenta of the $A-1$ and $A$ nuclear wave functions $\ket{\Psi^{J_{A-1}}_{A-1}}$ and $\ket{\Psi^{J_A}_A}$, respectively, and $u_{n}^{(\ell j)}(r)$ are the radial functions of the one-body Berggren basis states. {Note that Eq.~(\ref{E3}) is formulated in a translationally invariant manner in the COSM framework. This guarantees the absence of c.m.\ excitations and ensures that the overlap function is correctly expressed~\cite{PhysRevC.85.064320, michel2021gamow}.}


The SF is defined as the norm of the overlap function:
\begin{equation}
\label{E4}
C^2S_{\ell j} = \int_0^{+\infty} O(r)_{\ell j}^2~dr.
\end{equation}
In the GSM, SFs are independent of the chosen s.p.\ basis~\cite{michel2021gamow}. Note that the overlap function and corresponding SF are inherently complex in the GSM~\cite{BERGGREN1968265, PhysRevC.75.031301}, with the imaginary part interpreted as an uncertainty~\cite{PhysRevC.67.054311, PhysRevC.75.031301, BERGGREN1968265}. However, only the real parts are considered in this study. 
{Further to verify the robustness of our results with respect to the interaction choice, we also performed calculations using the phenomenological Minnesota two-body interaction~\cite{THOMPSON197753}. The resulting SFs and overlap functions closely match those employing the FHT interaction, with variations in the $\langle ^7\mathrm{Li} | ^6\mathrm{He} + p \rangle$ SF on the order of only 0.01. This indicates that the uncertainty of the interaction used in our calculated SFs is negligible. This conclusion agrees with the SFs reported in NCSM calculations~\cite{PhysRevC.110.054301}.}

\subsection{The nuclear reaction model: coupled channels Born approximation}

The DWBA was used in the original analysis of the \(^{6}{\rm He}(d,n)^7{\rm Li}\) data~\cite{Li2010} to extract SFs for the $\rm \langle ^7Li|^6He+p \rangle$ overlaps using conventional s.p.\ overlap functions. However, coupling between the $3/2^-$ ground and 0.478-MeV $1/2^-$ first excited states of $^7$Li, as well as ground state reorientation, can play a significant role in the reaction mechanism. Therefore, in this work we employ the coupled channels Born approximation (CCBA) where the transfer steps are described by the DWBA, but the inelastic couplings in the exit partition are explicitly included using the coupled channel (CC) formalism. All reaction calculations presented in this work were performed with the code {\sc fresco} \cite{THOMPSON1988167} and used the post form of the DWBA formalism with full complex remnant term.

\section{Results}

\subsection{Overlap functions and single particle wave functions\label{sec:OF}}

\begin{figure}[t]
    \centering
    \includegraphics[width=0.9\linewidth]{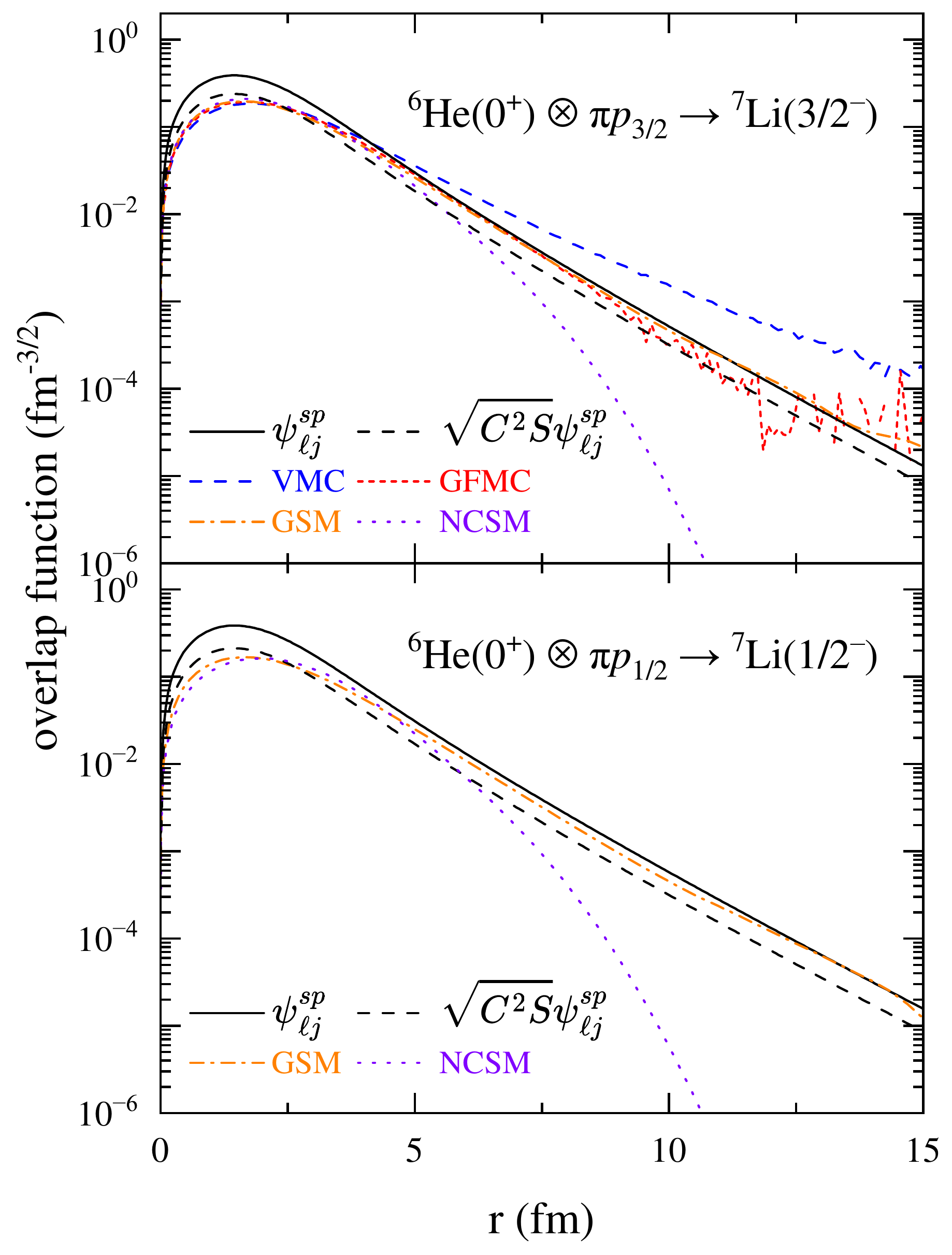}
    \caption{
    Comparison of the \(\rm \langle ^7Li|^6He+p \rangle\) overlap functions calculated with s.p.\ wave functions and those from GSM, NCSM, VMC, and GFMC calculations. $\sqrt{C^2S} \psi_{\ell j}^{sp}(r)$ denotes the s.p.\ wave function normalized by the GSM SF.}
    \label{overlap_function2}
\end{figure}

We first compare the s.p.\ wave functions for $\rm \langle ^7Li|^6He+p \rangle$, obtained using a WS potential well of ``standard'' geometry ($r_0 = 1.25$ fm, $a_0 = 0.65$ fm), with overlap functions calculated using various nuclear structure models, including the GSM, \textit{ab initio} no-nore shell model (NCSM)~\cite{PhysRevC.110.054301}, Variational Monte Carlo (VMC)~\cite{PhysRevC.84.024319}, and Green's Function Monte Carlo (GFMC) ~\cite{PhysRevC.84.024319}, as shown in Fig.~\ref{overlap_function2}. The VMC and GFMC results, taken from Ref.~\cite{WiringaOverlaps}, are available only for the $^7$Li ground 
state. As defined in Eq.~(\ref{E3}), the GSM and NCSM~\cite{PhysRevC.110.054301} overlap functions are plotted after dividing by $r$. The NCSM results were obtained used the Daejeon16 interaction~\cite{SHIROKOV201687} with $\hbar\omega = 15$ MeV and $N_{\rm max}=10$. 

The amplitude of the s.p.\ wave functions is notably larger than those of the theoretical models, primarily due to normalization differences. In structure models, overlap functions are scaled by their calculated SF~\cite{PhysRevC.75.031301, Michel_Springer, PhysRevC.84.024319, PhysRevC.110.054301}, as in Eq.~(\ref{E4}). For the weakly bound nucleus $^6$He, correlations and continuum coupling lead to SFs below unity. By contrast, s.p.\ wave functions obtained from WS potentials are unit-normalized, and the corresponding overlap function is written as~\cite{AUMANN2021103847}
$O_{\ell j}(r) \sim \sqrt{C^2S} \psi_{\ell j}^{sp}(r)$,
where $\psi_{\ell j}^{sp}(r)$ is the s.p.\ wave function and $\sqrt{C^2S}$ the spectroscopic amplitude. The $\sqrt{C^2S} \psi_{\ell j}^{sp}(r)$ are also presented for comparison, taking $C^2S$ as the real parts of the GSM SFs.
Due to the absence of correlations, the overlap functions \(\sqrt{C^2S} \psi_{\ell j}^{sp}(r)\) typically display higher amplitudes around 2 fm and reduced values around 3-4 fm, compared to those derived from the GSM and other models.

Focusing on the asymptotic region, the overlap functions calculated using \textit{ab initio} NCSM deviate significantly from the WS s.p.\ wave functions due to the absence of continuum coupling. The VMC overlap function also shows significantly different behavior with a much slower fall-off as a function of $r$. While the overlap function obtained from \textit{ab initio} GFMC shows a similar asymptotic behavior to the s.p.\ wave function, it is accurate only up to about 6 fm, after which it exhibits noticeable oscillations and discontinuities with large statistical uncertainties~\cite{PhysRevC.84.024319}. The VMC overlap function also has significant statistical errors at large $r$. Consequently, neither VMC nor GFMC overlaps can be directly used in reaction calculations without smoothing procedures~\cite{PhysRevC.86.024315}. By contrast, the GSM overlap functions show better consistency with the s.p.\ wave functions in the asymptotic region. However, the s.p.\ overlap function $\sqrt{C^2S} \psi_{\ell j}^{sp}(r)$ has a markedly different shape, so if normalized to the asymptote of the GSM overlap function would be significantly larger at smaller $r$.

{Asymptotic normalization coefficients (ANCs)  were extracted by fitting the GSM overlap functions to Whittaker functions in the asymptotic region, as described in Ref.~\cite{PhysRevC.85.064320}. Performing this fit at $r = 10$ fm, we extract ANC values of 2.587 fm$^{-1/2}$ and 2.361 fm$^{-1/2}$ for the ground and first excited state of $^7$Li. These values agree with the experimentally extracted average ANCs of 2.400 and 2.040 fm$^{-1/2}$ from the $^6$He$(d,n)^7$Li transfer reaction~\cite{doi:10.1142/S0218301323500350}. They are also close to the predictions of the shell model (SM), which are 2.377 and 2.020 fm$^{-1/2}$~\cite{PhysRevC.88.044315}. In contrast, the VMC calculations yield significantly larger ANC values of 3.68(5) and 3.49(5) fm$^{-1/2}$~\cite{PhysRevC.83.041001}, suggesting potential differences in the treatment of asymptotic behavior.}
{Note that Ref.~\cite{PhysRevC.83.041001} uses an integral method to calculate the ANC, which avoids the difficulties associated with obtaining reliable asymptotic behavior of variational wave functions.}

Due to the inter-nucleon correlations and continuum coupling being well treated within GSM calculations, the intra-nucleus and asymptotic behavior of the overlap functions are well described, providing more realistic inputs for transfer reaction calculations. 
However, the overlap function derived from the GSM exhibits noticeable oscillations beyond approximately 15 fm, attributed to precision loss in the numerical calculations. This can be mitigated by increasing the density of discretized continuum states in the Berggren basis. If required, in practical calculations this may be addressed by substituting the GSM overlap function with an s.p.\ wave function for $r > 15$ fm. 

\subsection{$^6\rm{He}($$d,n$$)^7\rm{Li}$ transfer reaction calculations}

The $^6$He($d,n$)$^7$Li  transfer reaction provides a test case to investigate the role of the overlap function in reaction calculations. An angular distribution was measured in inverse kinematics at an incident energy of $E_{\rm ^6He} = 36.4$ MeV by Li {\it et al.}~\cite{Li2010} using the HI-13 tandem accelerator in Beijing. The 0.0-MeV $3/2^-$ and 0.478-MeV $1/2^-$ levels were not separated in the experiment, so the data correspond to their total cross section.

\begin{figure}[t]
    \centering
    \includegraphics[width=0.9\linewidth]{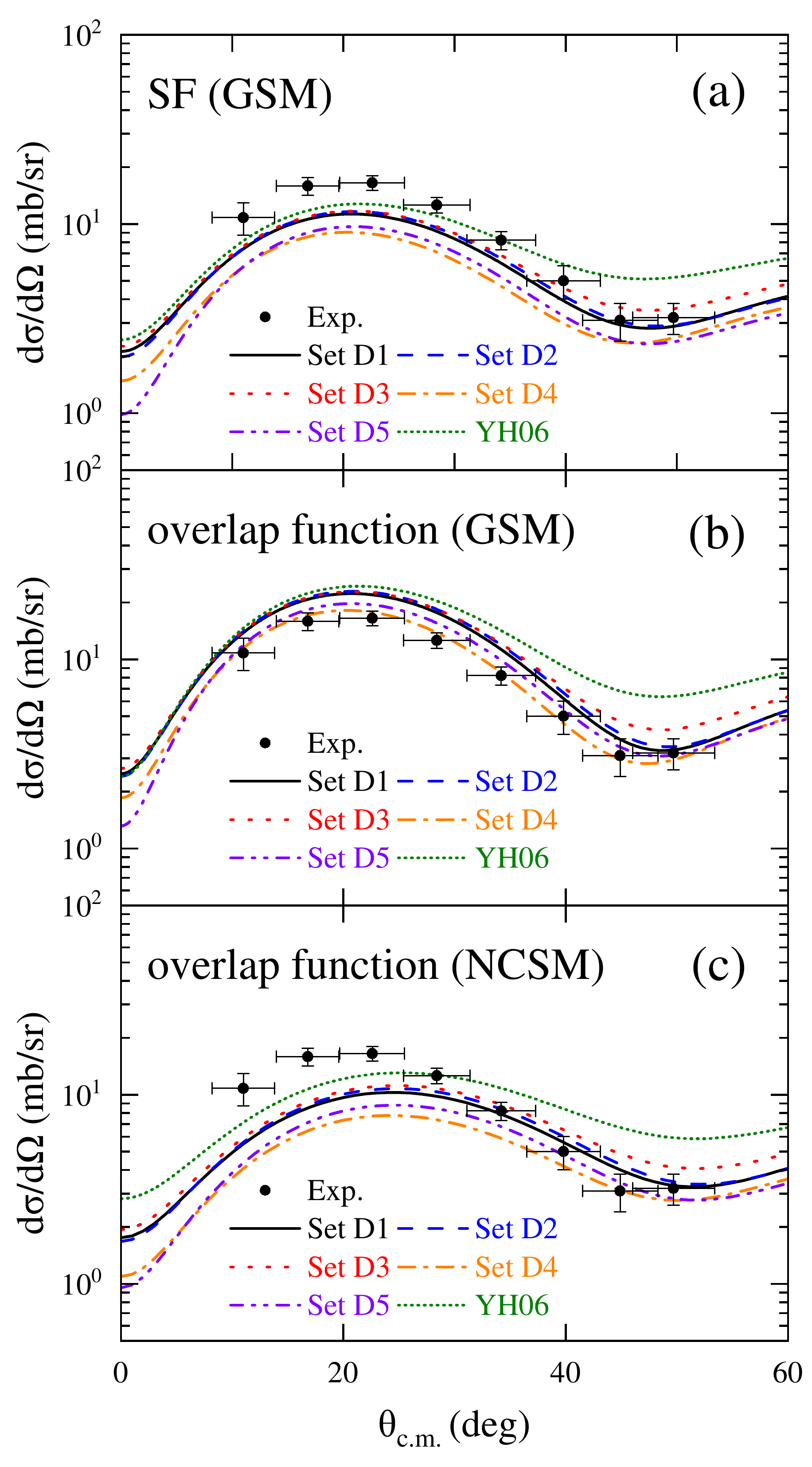}
    \caption{Comparison of angular distributions for the $^6$He($d,n$)$^7$Li transfer reaction calculated with different nuclear structure inputs for the $\rm \langle ^7Li|^6He+p \rangle$ overlap functions: (a) CCBA calculations using GSM SFs and s.p.\ wave functions, (b) CCBA calculations using overlap functions derived from the GSM, and (c) CCBA calculations using overlap functions derived from the NCSM.}
    \label{result}
\end{figure}

The same five sets of optical potential parameters were utilized in the entrance channel ($d$+$^6$He) as in the analysis of Ref.~\cite{Li2010}: D1, D2, D3, D4, and D5 from Refs.~\cite{PhysRevC.73.054605, PhysRevC.21.2253, HINTERBERGER1968265, VERNOTTE19941, PEREY19761}, respectively, plus the YH06 parameters from Ref.~\cite{PhysRevC.74.044615}. The CH89 global optical potential~\cite{VARNER199157} was employed in the exit channel ($n$+$^7$Li). Coupling between the ground and first excited states of $^7$Li was included using standard rotational model form factors and a deformation length of $\delta_2=2$ fm. 
The optical potential parameters were not adjusted after including the couplings since the predicted elastic scattering at the appropriate energy was almost identical to the no coupling result. 
The $\langle d|n+p \rangle$ overlap was calculated using the Reid soft core potential \cite{REID1968411}. Two different approaches were adopted for the $\rm \langle ^7Li|^6He+p \rangle$ overlap functions. 
For comparison purposes, we performed calculations using the traditional treatment based on Eq.~(\ref{E1}), with GSM-derived SFs and WS potentials of ``standard'' geometry, $r_0 = 1.25$ fm and $a_0 = 0.65$ fm. The results are plotted in Fig.~\ref{result}(a). In the second set of calculations the $\rm \langle ^7Li|^6He+p \rangle$ overlap functions were directly derived from the GSM, without relying on the s.p.\ approximation. The results are presented in Fig.~\ref{result}(b).

\begin{figure}
    \centering
    \includegraphics[width=0.9\linewidth]{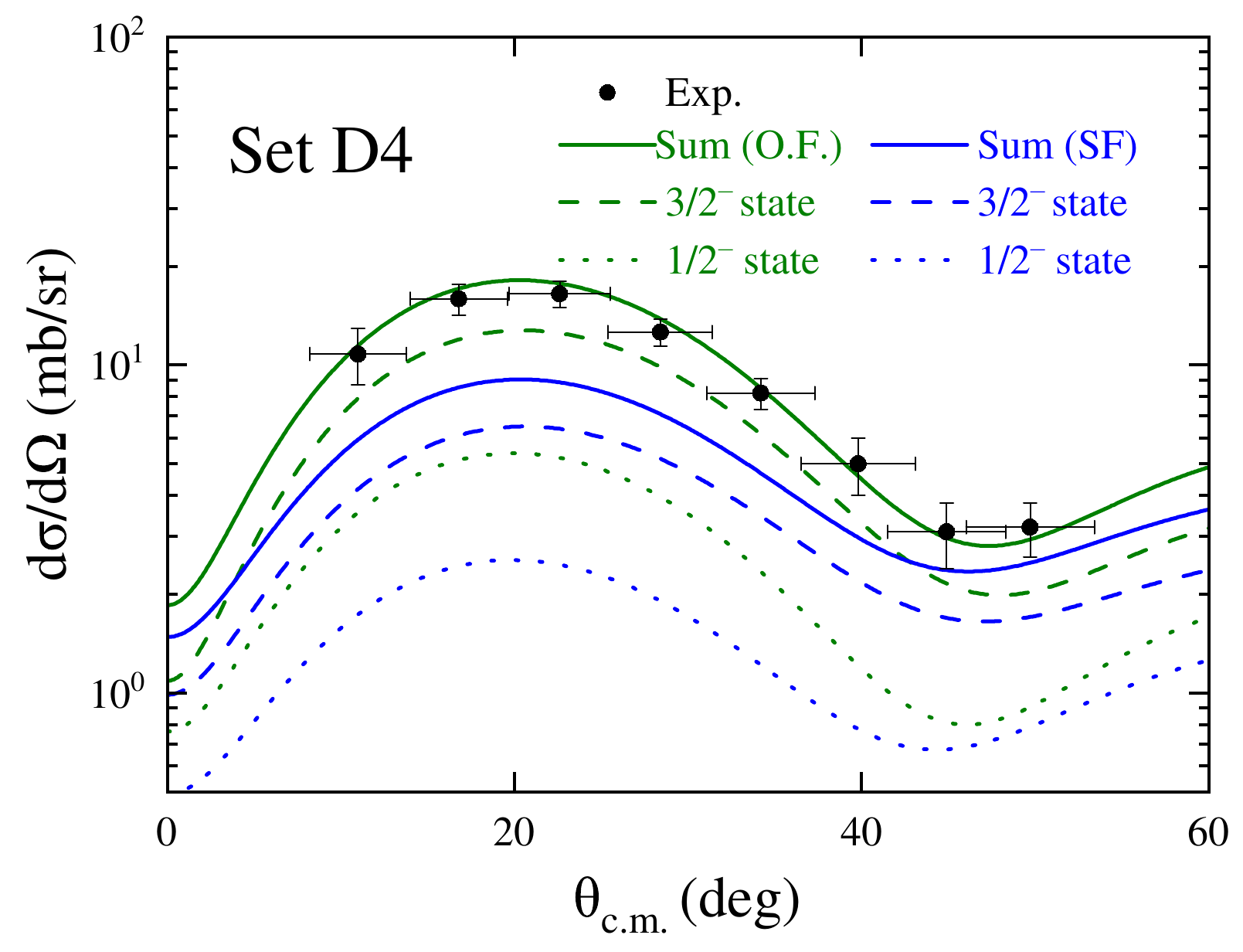}
    \caption{Angular distributions of the $^6$He($d,n$)$^7$Li reaction leading to the 0.0-MeV $3/2^-$ and 0.478-MeV $1/2^-$ states of $^7$Li, plus their sum, calculated using the D4 deuteron optical potential and GSM overlap functions (O.F.) and s.p.\ wave functions normalized by GSM SFs (SF), compared with the data of Ref.~\cite{Li2010}.}
    \label{parital}
\end{figure}

The results demonstrate that while calculations based on the s.p.\ wave functions and GSM SFs, which have correct asymptotic behavior, reproduce the peak position of the experimental angular distribution, they significantly underestimate its magnitude and fail fully to capture its shape.
The angular distributions calculated directly using the GSM overlap functions give a much better description of the data, not only in terms of its magnitude but also its shape, highlighting the importance of the internal structure of the overlap function in calculating reaction angular distributions.
Among the optical potentials tested, D4 combined with the GSM overlap functions gives the best description, matching the data within experimental uncertainties.
Fig.~\ref{parital} shows the decomposition of the total cross section calculated with D4 into $^7$Li $3/2^-$ and $1/2^-$ components using both GSM and s.p.\ overlap functions.

{Further to evaluate the role of the overlap functions, we performed additional calculations using \textit{ab initio} NCSM overlap functions, as shown in Fig.~\ref{result}(c). Although these functions lack correct asymptotic behavior, the resulting angular distributions are generally similar to those obtained using the s.p.\ overlap functions, which have proper asymptotics. This similarity highlights the importance of incorporating realistic nuclear structure information in the overlap functions.}
{It is worth noting that the NCSM with continuum (NCSMC), an extension of the NCSM, provides a unified microscopic framework for describing both bound and scattering states in light nuclei~\cite{PhysRevLett.110.022505, PhysRevC.103.035801, PhysRevC.87.034326}. The correct asymptotic behavior of wave functions for light nuclei such as $^6$Li, $^7$Be, and $^7$Li can be properly described within the framework of the NCSMC~\cite{Navrátil_2016, PhysRevC.100.024304}.}


\subsection{$^{208}\rm Pb ($$^7\rm Li$,$^6 \rm He$$) ^{209} \rm Bi$ transfer reaction calculations}

\begin{figure}[t]
    \centering
    \includegraphics[width=1\linewidth]{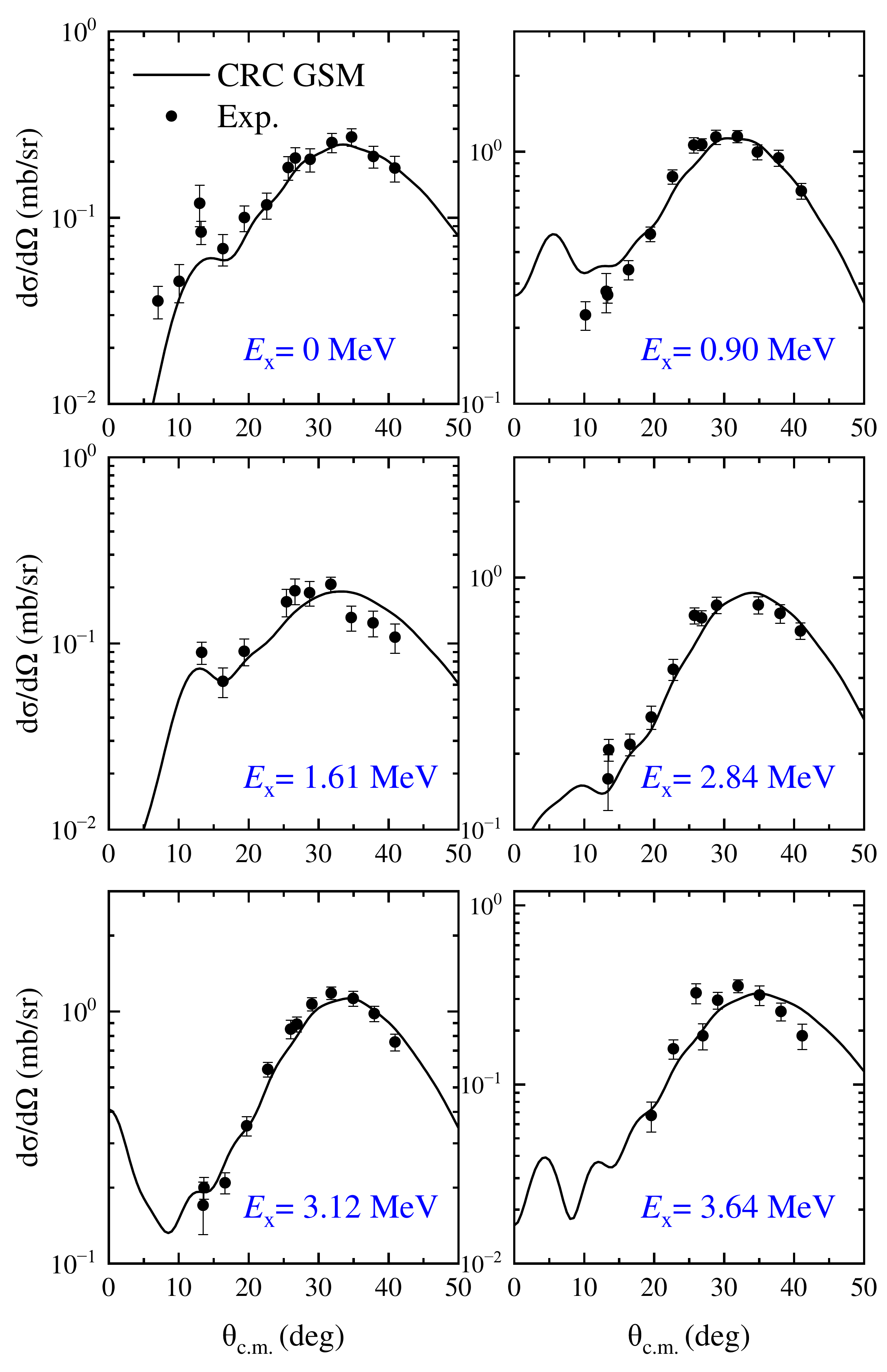}
    \caption{The 52-MeV $^{208}$Pb($^7$Li,$^6$He)$^{209}$Bi data of Zeller {\it et al.}~\cite{Zel79} (filled circles) compared with the results of CRC calculations employing the GSM $\left< ^7\mathrm{Li} \mid \protect{^6\mathrm{He}} + p \right>$ overlaps (solid curves).}
    \label{fig:zeller}
\end{figure}

The entrance channel of the $^6$He($d$,$n$)$^7$Li transfer reaction is an example of a system where both reaction partners are weakly bound and their simultaneous breakup could have a significant impact on the reaction mechanism. A complete treatment of breakup in such systems would require a four-body extension of the continuum-discretized coupled-channels (CDCC) method such as that recently proposed by Descouvemont \cite{Des18}. The weakly bound $^7$Li in the exit partition potentially further complicates the reaction mechanism. Due to the relatively low incident energy our analysis assumed that any breakup effects can be adequately represented by appropriate optical potentials, with the explicit inclusion of ground state reorientation and excitation of the low-lying bound $1/2^-_1$ state in $^7$Li taking care of the main couplings in the exit partition. However, the lack of suitable elastic scattering data introduces further uncertainties into the calculations due to the necessity of employing global potentials. Therefore, as a further check on the GSM overlap functions we analyzed the data of Zeller {\it et al.}~\cite{Zel79} for the $^{208}$Pb($^7$Li,$^6$He)$^{209}$Bi transfer reaction. It has been demonstrated that the CCBA provides a satisfactory description of the reaction mechanism for this reaction \cite{Kee20}. The calculations were similar to those described in Ref.~\cite{Kee20} with the s.p.\ $\left< ^7\mathrm{Li} \mid \protect{^6\mathrm{He}} + p \right>$ overlap functions replaced by the GSM. The full coupled reaction channels (CRC) formalism was also employed rather than the CCBA since this gave a slightly improved description of the data at the most forward angles for some of the populated $^{209}$Bi levels. The results are compared with the data in Fig.~\ref{fig:zeller} and show that the GSM overlap functions provide a good description of the data, supporting our findings based on the $^6$He($d$,$n$)$^7$Li data.

\subsection{Spectroscopic factors and ANCs}

While the shape of the $^6$He($d,n$)$^7$Li angular distribution is well reproduced by the CCBA calculations using the GSM overlap functions, the magnitude of the cross section is over predicted in most cases. Conversely, the calculations using the conventional s.p.\ approximation (Eq.~(\ref{E1})) with the GSM SFs under predict the data. To quantify this, ``experimental'' SFs and ANCs were extracted by normalizing the CCBA calculations to the data.
Since the measured angular distribution \cite{Li2010} represents the sum of the cross sections populating the 0.0-MeV $3/2^-$ and 0.478-MeV $1/2^-$ levels of $^7$Li and the calculated angular distributions for these levels have the same shape, see Fig.~\ref{parital}, it is not possible to distinguish the individual contributions from the transferred proton in the \(p_{3/2}\) and \(p_{1/2}\) orbitals.
Therefore,  we fixed the ratio of SFs for populating the ground and excited states of \(^7\text{Li}\) to the GSM value. 

\begin{figure}
    \centering
    \includegraphics[width=0.9\linewidth]{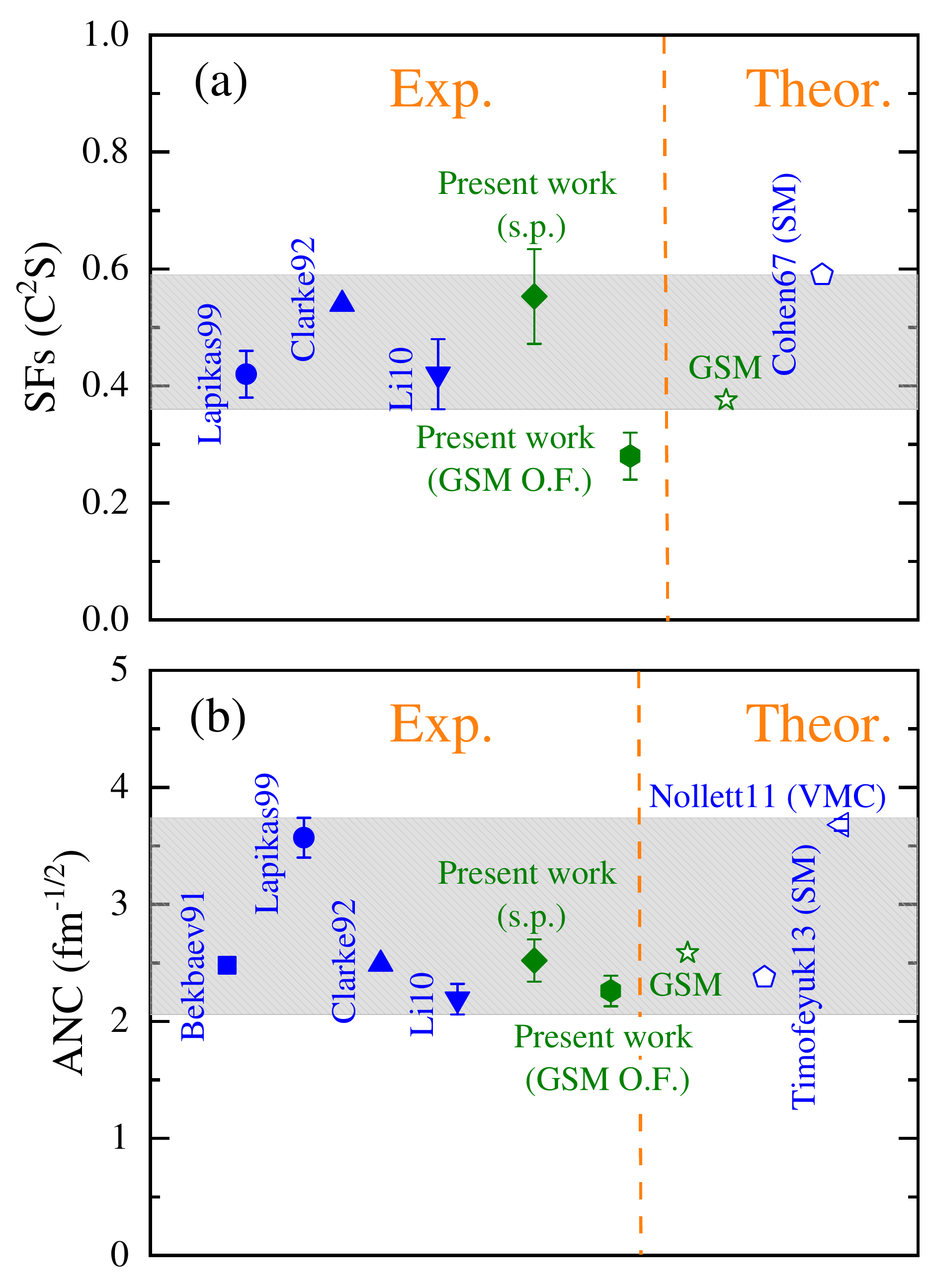}
    \caption{Comparison of SFs (a) and ANCs (b) for the ground state of \( ^7 \mathrm{Li} \) extracted from the \(^{6}\text{He}(d,n)^7\text{Li}\) data by the present CCBA analyses using s.p. wave functions (s.p.) and GSM overlap functions (GSM O.F.) with SFs from other work: Li10~\cite{Li2010}, Clarke92~\cite{Clarke_1992}, Lapikas99~\cite{PhysRevLett.82.4404} and Bekbaev91~\cite{Bek91} (solid blue symbols), and theoretical values (open symbols) calculated using the GSM, conventional SM~\cite{PhysRevC.88.044315,COHEN19671}, and VMC~\cite{PhysRevC.83.041001}.
    }
    \label{SF}
\end{figure}

Fig.~\ref{SF}(a) compares the single-proton SFs for the ground state of $^7$Li deduced from the \( ^7{\rm Li}(t,\alpha)^6{\rm He}\)~\cite{Clarke_1992} and \(^7{\rm Li}(e,e'p)\)~\cite{PhysRevLett.82.4404} reactions, plus the DWBA analysis of the \( ^6{\rm He}(d,n)^7{\rm Li}\) data by Li {\it et al.}~\cite{Li2010} with the values obtained in the present work using this procedure. These are represented by the mean of the results for different deuteron optical potentials, with the uncertainty given by the standard deviation. 
Theoretical SFs from the GSM and standard SM~\cite{COHEN19671} are also plotted.
{Comparison of SFs obtained from fits to reaction data using the s.p.\ approximation can be problematical, due to their well known dependence on binding potential geometry. The SFs obtained in the present work (s.p.) and by Clarke \cite{Clarke_1992} (Clarke92) and Li {\it et al.}~\cite{Li2010} (Li10) employ the same geometry.}
{The present result agrees well with Clarke \cite{Clarke_1992}, while that of Li {\it et al.}~\cite{Li2010} is significantly smaller, just agreeing within the estimated uncertainties. Since the present work and Li {\it et al.}~\cite{Li2010} analyze the same data set with different reaction models, CCBA and DWBA, the difference underscores the reaction model dependence also inherent in SF extractions. The present work additionally uses a somewhat more realistic $\rm \langle d|n+p \rangle$ overlap.}




{The \(^7{\rm Li}(e,e'p)\) analysis of Lapik\'as {\it et al.}~\cite{PhysRevLett.82.4404} uses different binding potential parameters. The rms radius of the s.p.\ wave function quoted in Ref.~\cite{PhysRevLett.82.4404} corresponds to a WS binding potential with $r_0 = 1.90$ fm, significantly larger than the ``standard'' value of 1.25, if $a = 0.65$ fm. The apparent good agreement between the SF and that obtained by Li {\it et al.}~\cite{Li2010} may therefore be fortuitous. The SF obtained by re-normalizing the GSM overlap (GSM O.F.) is significantly smaller than the other ``experimental'' values. Of the theoretical values, the standard SM agrees well with the present s.p.\ result and Clarke \cite{Clarke_1992} while the GSM is a good match to the results of Lapik\'as {\it et al.}~\cite{PhysRevLett.82.4404} and Li {\it et al.}~\cite{Li2010}.}

\begin{figure}[t]
    \centering
    \includegraphics[width=0.9\linewidth]{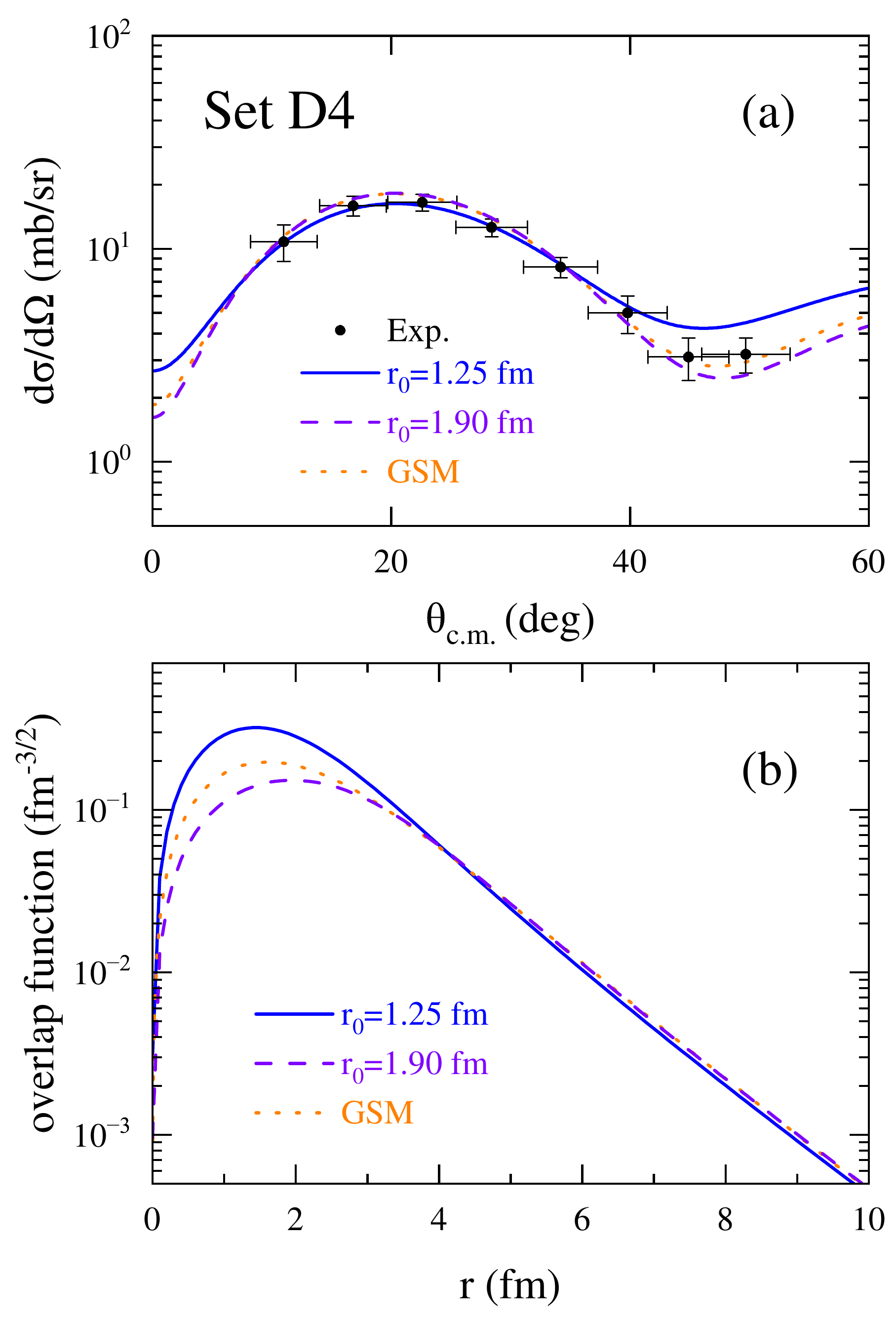}
    \caption{(a) CCBA fits to the measured \(^{6}\text{He}(d,n)^7\text{Li}\) angular distribution \cite{Li2010} with s.p.\ overlap functions calculated from WS potentials with $r_0 = 1.25$ fm and $1.90$ fm, plus the result using the GSM overlap. All calculations used the D4 deuteron optical potential in the entrance channel. (b) The corresponding $\rm \langle ^7Li|^6He+p \rangle$ overlaps for the $^7$Li $3/2^-$ ground state, normalized by the relevant SFs for the s.p.\ wave functions.} 
    \label{r0}
\end{figure}

{In Fig.~\ref{r0}(a), we compare CCBA fits to the \( ^6{\rm He}(d,n)^7{\rm Li}\) angular distribution using the D4 deuteron optical potential and s.p.\ $\rm \langle ^7Li|^6He+p \rangle$ overlaps calculated with WS potentials of $r_0 = 1.90$ and 1.25 fm, plus the result with the GSM overlap functions, as in Fig.~\ref{parital}. The calculations employing the s.p.\ wave functions are normalized to the data using SF values of 0.678 and 0.541 for $p_{3/2}$ and $p_{1/2}$, respectively with $r_0 = 1.25$ fm, and 0.308 and 0.246 with $r_0 = 1.90$ fm. The value of 0.308 for the $p_{3/2}$ is somewhat smaller than the value of 0.42(4) extracted from the \(^7{\rm Li}(e,e'p)\) data by Lapik\'as.}

The s.p.\ result with $r_0 = 1.90$ fm is very close to that with the GSM overlap functions, while the result for $r_0 = 1.25$ fm gives a poorer description of the stripping peak shape. 
Fig.~\ref{r0}(b) plots the corresponding $\rm \langle ^7Li|^6He+p \rangle$ overlaps for the $^7$Li $3/2^-$ ground state, normalized by the relevant SFs. 
All three overlaps have similar asymptotic behavior, the small differences for radii $r > 5$ fm are within the range permitted by the experimental uncertainties in the data. However, the s.p.\ overlap for $r_0 = 1.90$ fm continues closely to match the GSM result for $r > 3$ fm, whereas the overlap for $r_0 = 1.25$ fm increasingly deviates from it as the radius is reduced from 5 fm. 
The difference in the shape of the angular distribution for the calculation using the s.p.\ wave function with $r_0 = 1.25$ fm and those of the calculations with $r_0 = 1.90$ fm and the GSM overlaps is thus presumably linked to the shape of the overlap in the range $3 < r < 5$ fm.

{Fig.~\ref{SF}(b) plots single proton ANCs for the ground state of $^7$Li obtained by fitting data (filled symbols) or from different theoretical calculations (open symbols). Comparison of ANCs should be more straightforward, since if the reaction is truly peripheral the analysis will only be sensitive to the asymptote of the overlap functions, i.e.\ the ANC, which only weakly depends on the geometry of the binding potential in the s.p.\ approximation. There is excellent agreement among the experimental and theoretical ANCs, with the exception of the value extracted from the fit of Lapik\'as {\it et al.} to the \(^7{\rm Li}(e,e'p)\) data~\cite{PhysRevLett.82.4404} and the VMC result \cite{PhysRevC.83.041001} which, however, agree well with each other. Note that if the $^6$He($d$,$n$)$^7$Li data are fitted with the same s.p.\ overlap function as Lapik\'as {\it et al.}~\cite{PhysRevLett.82.4404} a mean ANC of $2.697 \pm 0.203$ fm$^{-1}$ is obtained, much closer to the other values. This probably reflects the ability of the \((e,e'p)\) reaction to probe the overlap function over the whole radial range, whereas typically direct reactions only probe the exponential tail~\cite{Kra01}.}

These results suggest that more precisely to deduce the SFs from experimental cross sections, a more accurate approach should be employed. For instance, begin with an overlap function obtained from a state-of-the-art nuclear structure model as the input for the transfer reaction calculations. This will provide a more realistic description of the nuclear wave function than the s.p.\ approximation.
In the subsequent transfer reaction calculations, a quenching factor $Q$ is then introduced to scale the calculated cross section to match the experimental data, following the relation \((d\sigma/d\Omega)_{\rm exp} = Q\cdot (d\sigma/d\Omega  )_{\rm th}\). 
The experimental SF is then obtained as \(C^2S_{\rm exp} = Q \cdot C^2S_{\rm th}\), ensuring that the extracted SFs reflect both the theoretical input and the experimental observations.
Applying this procedure to the present case yields a mean quenching factor $Q = 0.78 \pm 0.09$, derived from six deuteron optical potentials, resulting in SFs of $0.29 \pm 0.04$ and $0.23 \pm 0.03$ for the $^7$Li ground and $1/2^-$ first excited states, respectively. 
{The uncertainties in the cross section data add about 5\% to the uncertainty in $Q$. The resulting SF for the ground state is significantly smaller than most of the other determinations, although the good agreement of the corresponding ANCs demonstrates that the asymptotic behavior of the GSM overlap is correct. The SF reflects the importance of the inter-nucleon correlations in reducing the magnitude of the GSM overlap compared to the normalized s.p.\ wave function at smaller radii, see Fig.~\ref{overlap_function2}.}

{This result is also consistent with the long-standing discrepancy between theoretical and experimental SFs which remains a major challenge in nuclear reaction studies~\cite{PhysRevC.77.044306, PhysRevC.92.041302, PhysRevLett.131.212503, AUMANN2021103847, PhysRevLett.104.112701, PhysRevLett.107.032501, PhysRevC.83.034610, PhysRevC.67.064301, PhysRevC.97.034601, PhysRevC.73.044608}, leading to a non-unit quenching factor. A fully consistent calculation of both bound and scattering aspects of the reaction should further improve the results. For example, use of the dispersive optical model in knockout reactions to provide a consistent treatment of both bound and scattering states, thereby achieving fully self-consistent reaction calculations, gives a good description of the $^{40}$Ca$(e,e'p)^{39}$K reaction~\cite{PhysRevC.98.044627}.}



\section{SUMMARY}

This study examines the role of the overlap function in single-nucleon transfer reactions, focusing on the \(^{6}{\rm He}(d,n)^7{\rm Li}\) reaction. Employing spectroscopic factors and overlap functions from the Gamow shell model, we calculated angular distributions and compared them with experimental data from Ref.~\cite{Li2010}. Our results indicate that the GSM-derived overlap function accurately captures both the asymptotic behavior and internal characteristics of the wave function, successfully reproducing the observed angular distribution.
However, transfer calculations using the single particle approximation yield results that differ significantly from those obtained using GSM overlap functions.
This discrepancy underscores the importance of the internal structure of the overlap function in transfer reaction calculations. 
Additionally, calculations with overlap functions from the no-core shell model, lacking continuum coupling, do not capture the asymptotic behavior and produce notably different angular distributions. 
{We note, however, that the no-core shell model with continuum is able to describe correctly the asymptotic behavior of wave functions for light nuclei such as $^{6,7}$Li and $^7$Be \cite{Navrátil_2016, PhysRevC.100.024304}.}
These comparisons demonstrate the necessity of incorporating detailed nuclear structure effects, particularly those involving continuum coupling and inter-nucleon correlations, for precise modeling of nuclear reactions.

\textit{Acknowledgments.}~
This work was supported by the National Key R\&D Program of China under Grant Nos.\ 2024YFE0109800, 2024YFE0109802, and 2023YFA1606403; the National Natural Science Foundation of China under Grant Nos.\  12205340, 12175281, and 12121005;  the Gansu Natural Science Foundation under Grant No.\ 25JRRA467;  the Strategic Priority Research Program of the Chinese Academy of Sciences under Grant No.\ XDB34000000; the Key Research Program of the Chinese Academy of Sciences under Grant No.\ XDPB15; the State Key Laboratory of Nuclear Physics and Technology, Peking University under Grant No.\ NPT2020KFY13. The numerical calculations in this paper were performed at the Hefei advanced computing center.

\section*{References}

\bibliographystyle{elsarticle-num_noURL}
\bibliography{Ref}

\end{document}